\documentclass{aa}

\usepackage{txfonts}
\begin{document}

   \title{Molecular jet emission and a spectroscopic survey of S235AB}

   \subtitle{}

   \author{R. A. Burns
          \inst{1}
          \and
          T. Handa\inst{1}\fnmsep\
          \and
          T. Hirota\inst{2,3}\fnmsep\
          \and
          K. Motogi\inst{4}\fnmsep\
          \and
          I. Imai\inst{1}\fnmsep\
          \and
          T. Omodaka\inst{1}\fnmsep\
          }

   \institute{Graduate School of Science and Engineering, Kagoshima University, 1-21-35 K\^orimoto, Kagoshima 890-0065, Japan\\
              \email{RossBurns88@Googlemail.com}
         \and
           Mizusawa VLBI Observatory, National Astronomical Observatory of Japan, Osawa 2-21-1, Mitaka, Tokyo 181-8588, Japan
         \and
           Department of Astronomical Sciences, SOKENDAI (Graduate University for Advanced Studies), Osawa 2-21-1, Mitaka, Tokyo 181-8588, Japan
         \and
           Mizusawa VLBI Observatory, National Astronomical Observatory of Japan, 2-12 Hoshi-ga-oka, Mizusawa-ku, Oshu-shi, Iwate 023-0861, Japan
             }

   \date{Received August 24, 2015; accepted December 12, 2015}

% \abstract{}{}{}{}{} 
% 5 {} token are mandatory
 
  \abstract
   {The S235AB star forming region houses a massive young stellar object which has recently been reported to exhibit possible evidence of jet rotation - an illusive yet crucial component of disk aided star formation theories.}
   {To confirm the presence of a molecular counterpart to the jet and to further study the molecular environment in in S235AB.}
   {We search for velocity wings in the line emission of thermal SiO (J=2-1, v=0), a tracer of shocked gas, which would indicate the presence of jet activity. Utilising other lines detected in our survey we use the relative intensities of intra species transitions, isotopes and hyperfine transitions to derive opacities, temperatures, column densities and abundances of various molecular species in S235AB.}
   {The SiO (J=2-1, v=0) emission exhibits velocity wing of up to 75 km s$^{-1}$ above and below the velocity of the star, indicating the presence of a jet. The molecular environment describes an evolutionary stage resemblant of a hot molecular core.}
   {}


   \maketitle

\section{Introduction}

The target of our study, S235AB, is a star forming region that houses a massive young stellar object (MYSO) called S235AB-MIR - which was recently reported to exhibit fast ($\sim 50$ km s$^{-1}$) jet-tracing water masers with a signature of rotation \citep{Burns15a}. 
The confirmation of a rotating jet in an MYSO would have great impact on theories of star formation as it would provide a method of removing angular momentum from the inner disk of the system - enabling accretion - while also corroborating a magneto-centrifugal launching mechanism for jets from massive young stars (\emph{see} \citealt{Konigl00}). 

%Accretion rates in massive young stars are tightly correlated with outflow rates \citep{Garatti15}. As such, the nature of jets in MYSOs, particularly those exhibiting rotation, should be of special interest and importance to star formation theories and observation.

S235AB-MIR is known to have slow molecular outflows \citep{Felli04}, however, aside from the fast water maser velocities there is no evidence in the literature of a \emph{fast} molecular jet in S235AB-MIR (\citealt{Felli04} claimed to have found a continuum jet aligned with the water maser jet, however they later retracted the claim in \citealt{Felli06}).
As a follow-up to the maser jet results of \citet{Burns15a} the primary aim of our spectral line survey was to find the molecular counterpart to the water maser jet in S235AB, which should be readily identifiable from doppler broadened line `wings' seen in the thermal gas jet tracer SiO (J=2-1, v=0) (\emph{for example} \citealt{Ana11}). The secondary aim of our observations was to further investigate the physical gas conditions in S235AB via basic astrochemical study using auxiliary spectral lines.

S235AB is known to have a dense molecular core centered on S235AB-MIR and molecular outflows mapped in HCO$^+$ (1-0) and C$^{34}$S (5-4) \citep{Felli04}, CS (7-6) \citep{Wu10} and $^{13}$CO \citep{Felli97}. The \emph{Spitzer} colours indicate that S235AB-MIR has a mass of 11 M$_{\odot}$, making it the only MYSO in the region \citep{Dew11}, and non-detection of centimeter emission in S235AB attests to its youth \citep{Tofani95,Felli06}.

\begin{table*}[]
\caption{Summary of detected molecular lines and their measured parameters.\label{TAB}}
\begin{center}
\footnotesize
\begin{tabular}{cccccccc}
\hline
Molecular&Transition&$T_{A}^*$&Obs Frequency&$\Delta v$&$v_{peak}$&rms\\
formula &&[K] & [MHz] & [MHz] & [km/s] & [K] \\
\hline
%USB\\
%\hline
CH$_{3}$OH	 & 2(1,2)-1(1,1) A++	& 0.230 & 95914.406 & 0.819 & -17.22&0.012 \\
C$^{34}$S	 & 2-1			& 1.037 & 96413.016 & 0.852 & -17.13&0.008 \\
CH$_{3}$OH	 & 2(-1,2)-1(-1,1) E	& 0.962 & 96739.445 & 0.778 & -17.18&0.006 \\
CH$_{3}$OH	 & 2(0,2)-1(0,1) A++	& 1.330 & 96741.422 & 0.778 & -17.14&0.012 \\
CH$_{3}$OH	 & 2(0,2)-1(0,1) E	& 0.461 & 96744.617 & 0.836 & -17.21&0.016 \\
CH$_{3}$OH	 & 2(1,1)-1(1,0) E	& 0.210 & 96755.602 & 0.770 & -17.29&0.007 \\
C$^{33}$S	 & 2-1 3/2-3/2		& 0.030 & 97169.438 & 0.811 & -16.96&0.002 \\
C$^{33}$S	 & 2-1 1/2-1/2		& 0.151 & 97171.828 & 0.811 & -16.96&0.002 \\
C$^{33}$S	 & 2-1 3/2-1/2		& 0.038 & 97175.188 & 1.155 & -16.74&0.003 \\
OCS		 & 8-7			& 0.048 & 97301.203 & 0.786 & -16.98&0.005 \\
CH$_{3}$OH	 & 2(1,1)-1(1,0) A--	& 0.191 & 97582.859 & 0.803 & -17.16&0.005 \\
$^{34}$SO	 & N,J=2,3-1,2		& 0.094 & 97715.461 & 0.868 & -17.18&0.002 \\
CS		 & 2-1			& 3.639 & 97981.016 & 2.146 & -17.19&0.019 \\
CH$_{3}$CHO	 & 5(1,4)-4(1,3) E	& 0.030 & 98863.484 & 1.262 & -17.16&0.004 \\
SO		 & N,J=2,3-1,2		& 1.531 & 99299.953 & 0.901 & -17.14&0.013 \\
%\hline
%LSB\\
%\hline
CH$_{3}$OH     & 5(-1,5)-4(0,4) E& 2.471 & 84521.188 & 3.170 & -16.94&0.020 \\
OCS		 & 7-6		& 0.075 & 85139.250 & 0.606 & -17.51 &0.004 \\
HC$^{18}$O$^{+}$ & 1-0		& 0.035 & 85162.375 & 0.959 & -17.53 &0.005 \\
c-C$_{3}$H$_{2}$ & 2(1,2)-1(0,1)& 0.093 & 85338.969 & 0.614 & -17.22 &0.005 \\
HCS$^{+}$	 & 2-1		& 0.129 & 85347.859 & 0.721 & -16.96 &0.004 \\
CH$_{3}$CCH	 & 5(2)-4(2)	& 0.034 & 85450.820 & 0.311 & -17.19 &0.004 \\
CH$_{3}$CCH	 & 5(1)-4(1)	& 0.058 & 85455.727 & 0.688 & -17.22 &0.003 \\
CH$_{3}$CCH	 & 5(0)-4(0)	& 0.085 & 85457.219 & 0.287 & -16.72 &0.001 \\
HC$^{15}$N	 & 1-0		& 0.157 & 86054.992 & 0.795 & -17.09 &0.005 \\
SO		 & N,J=2,2-1,1	& 0.376 & 86093.984 & 0.721 & -17.00 &0.005 \\
H$^{13}$CN	 & 1-0 F=1-1	& 0.322 & 86338.750 & 0.786 & -17.05 &0.005 \\
H$^{13}$CN	 & 1-0 F=2-1	& 0.593 & 86340.180 & 0.811 & -17.01 &0.004 \\
H$^{13}$CN	 & 1-0 F=0-1	& 0.129 & 86342.242 & 0.688 & -16.95 &0.004 \\
HCO & 1(0,1)-0(0,0) 3/2-1/2 F=2-1& 0.044 & 86670.648 	 & 0.664 & -16.40&0.005 \\
H$^{13}$CO$^{+}$ & 1-0		& 0.282 & 86754.352 	 & 0.746 & -17.22&0.006 \\
SiO	 & 2-1 v=0	& 0.056 & 86847.023 	 &  1.901 & -17.04&0.001\\
%SiO (wings)	 & 2-1 v=0	& 0.011 & 86847.680 	 & 23.857 & -19.30&0.002\\
HN$^{13}$C	 & 1-0 F=2-1	& 0.124 & 87090.812 	 & 0.705 &-16.84& 0.005 \\
C$_{2}$H	 & 1-0 3/2-1/2 F=1-1& 0.137 & 87284.180 & 0.352 &-17.08& 0.003 \\
C$_{2}$H	 & 1-0 3/2-1/2 F=2-1& 0.890 & 87316.969 & 0.770 &-17.15& 0.008 \\
C$_{2}$H	 & 1-0 3/2-1/2 F=1-0& 0.479 & 87328.641 & 0.754 &-17.06& 0.006 \\
C$_{2}$H	 & 1-0 1/2-1/2 F=1-1& 0.637 & 87402.102 & 0.778 &-17.34& 0.008 \\
C$_{2}$H	 & 1-0 1/2-1/2 F=0-1& 0.296 & 87407.227 & 0.696 &-17.21& 0.010 \\
C$_{2}$H	 & 1-0 1/2-1/2 F=1-0& 0.149 & 87446.523 & 0.819 &-17.03& 0.008 \\
\hline
\end{tabular}

%\begin{tablenotes}
%\item $*$ Column densities which have been corrected for the effect of optical depth.
%\end{tablenotes}

\end{center}
\end{table*}

\section{Observations and Data Reduction}
We conducted molecular line observations of S235AB in March and April of 2015 with the 45-m radio telescope of the Nobeyama Radio Observatory (NRO), a branch of the National Astronomical Observatory of Japan. 

% 13th of March and 27th of April 2015 

We observed in two sideband mode using the TZ receiver and we observed horizontal and vertical linear polarisations simultaneously. The upper sideband (USB) and lower sideband (LSB) central frequencies were 85.85 and 97.85 GHz, respectively, with band widths of 4 GHz each. The beamsize was about 18$^{\prime\prime}$. The SAM45 spectrometer provided 16 independent frequency arrays (8 per polarisation) which we organised to provide continuous coverage of a wide frequency range, and to include lines of particular interest. The frequency resolution was 244.14 kHz, providing a bandwidth of 1000 MHz per array.

Observing coordinates were centered to the position of S235AB-MIR at 
$(\alpha, \delta)_{\mathrm{J}2000.0}=(05^{\mathrm{h}}40^{\mathrm{m}}53^{\mathrm{s}}.384$ +35$^{\circ}$41'48".447).
Sky-level subtraction was performed using a region of empty sky at coordinates $(\alpha, \delta)_{\mathrm{J}2000.0}=(05^{\mathrm{h}}40^{\mathrm{m}}58^{\mathrm{s}}.30$ +35$^{\circ}$41'48".60).
Absolute flux calibration was performed using the chopper wheel method, empty sky and hot load. Pointing accuracy was checked using a nearby SiO maser source, RU Aur, every $\sim 1-2$ hours.

During the March observations the atmospheric conditions were good, with typical $T_{sys} \approx$  140 K, however strong wind conditions resulted in poor pointing accuracy and caused much of the on-source scans to suffer from beam warping and pointing offsets. Only 1 hour of observing session provided usable data.
 
During the April observations wind conditions were much better, providing good pointing accuracy (usually better than 3$^{\prime\prime}$), and $T_{sys}$ was between 250 and 300 K. Data from the full 4 hour observing session was used.

To detect as many molecular lines as possible we interchanged between two frequency array setups which each covered different frequency ranges. The only frequency array common to both setups was that containing the SiO line - which was our transition of highest priority. Subsequently the frequency array containing the SiO line was observed for the full observing time while other frequency arrays were observed for roughly half of this time.

Data reduction was performed using the NEWSTAR software which was developed by the NRO. To reduce the aforementioned effect of the wind in the March observations we inspected all scans individually. Scans with bad pointing were determined by the failure in detecting C$^{34}$S, a rarefied gas whose region of peak emission is known to be compact from the maps of \citet{Felli04} (their Figure 9).
This greatly improved the quality of integrated spectra.

After flagging bad scans, 2$^{nd}$ order polynomial baselines were fit to individual scans. Finally, all scans were integrated and polarisations were merged. Good scans on the target integrated to a total of 1 hour for the March data and 2 hours for the April data. 
The achieved sensitivity was typically better than 0.006 K for most frequency arrays, and reached 0.0025 K for the array containing the SiO emission. The antenna gain was roughly 4.4 Jy/K.
%We converted to a main beam temperature scale using a beam efficiency of 40\%, as advertised on the NRO website.

\section{Results}

Gaussian profiles were fit to spectral lines detected above 5 times the r.m.s noise. The detections and line parameters are listed in Table~\ref{TAB}.
Detected molecular transitions were identified by cross-referencing with the Cologne Database for Molecular Spectroscopy (CDMS) catalogue \citep{CDMS}.

%USB and LSB spectra are shown in Figures~\ref{SP1}-\ref{SP2}.

%S235AB was found to be rich in molecular species, as is expected in the environment of a star forming region. There was very little deviation in the peak velocities of different species, with all lines typically between -16 to -18 km s$^{-1}$ with respect to the local standard of rest.

\subsection{SiO - jet tracer}
\begin{figure}[h!]

\begin{center}
\includegraphics[scale=0.44]{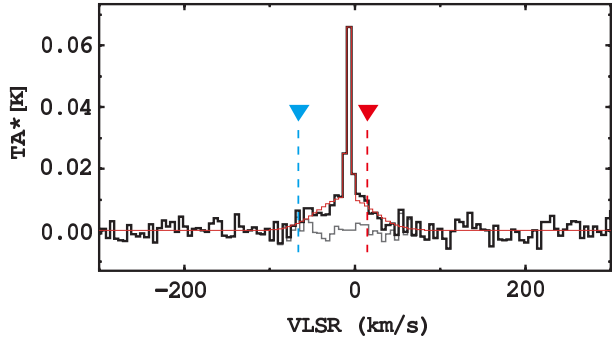}
\caption{Four channel binned spectrum of the SiO (2-1) emission in S235AB. Black lines show the observed data while the combined two Gaussian model and its residuals are shown in red and grey respectively. Blue and red triangles indicate extremum velocities of the H$_{2}$O masers of \citet{Burns15a}.
\label{SP3}}

\end{center}

\end{figure}

Our survey detected the common jet tracer transition line of SiO $(2-1)$. In low temperature environments SiO molecules are confined to dust grains, resulting in a low gas abundance. In a shocked gas environment these molecules are released from dust grains and become abundant in the gas phase \citep{Martin92}. As such SiO is a useful tracer of shocked gas and velocity wings in its spectrum indicate the presence of jets and outflows \citep{Ana11}. The spectrum of SiO $(2-1)$ gas in S235AB is displayed in Figure~\ref{SP3}, showing wide velocity wings.

%SO is seen as jet `bullets' in \citet{Zap10}.

%Although CS is not commonly associated with jets or outflows 
%\citet{Felli06} found a low velocity outflow via the extended blueshifted and redshifted lobes of CS gas in S235AB-MIR. The CS outflow manifests as a wide line width in our observations (Table~\ref{TAB}, column 5).

\subsection{Rotation Temperature}

We detected multiple rotational transitions of Sulphur monoxide (SO) such that it was possible to estimate its rotation temperature, $T_{rot}$, by using the `population diagram' method, detailed in \citet{Goldsmith99}. Emission was optically thin and the derived rotation temperature was $T_{rot}=26 \pm 4$ K which is similar to that obtained by \citet{Felli04} of $\sim 30$ K, using transitions of CH$_3$CN$(5-4)$.

%26.1 K

\begin{figure}[t]

\begin{center}
\includegraphics[scale=0.5]{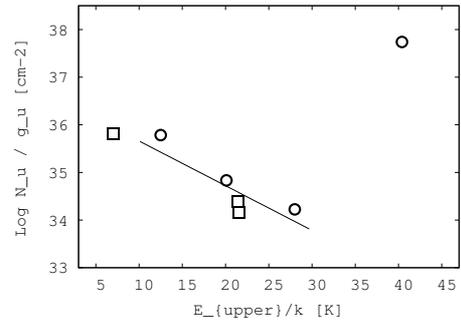}
\caption{Population diagram of CH$_3$OH in S235AB. E-and A-species are indicated by circles and squares respectively.
\label{POP}}
\end{center}
\end{figure}

We also measured the rotation temperature of methanol using transitions of CH$_3$OH (2-1), using lines of both E and A-species. The population diagram is shown in Figure~\ref{POP}, the slope giving T$_{rot}=10\pm1$ K, and the gas optically thin. The line fit ignores the CH$_3$OH (5-4) E transition at $E_{upper}=40$ K, which we discuss below. Methanol in this frequency regime is thought to be sub-thermally excited \citep{Menten88}. Our estimate may therefore be considered as the lower limit of the kinematic temperature of the methanol cloud. 

Multiple transitions of OCS were detected, but were too weak to attain a meaningful temperature estimate.

\subsection{Detection of a class I methanol maser}

\begin{figure}[t]
\begin{center}
\includegraphics[scale=0.43]{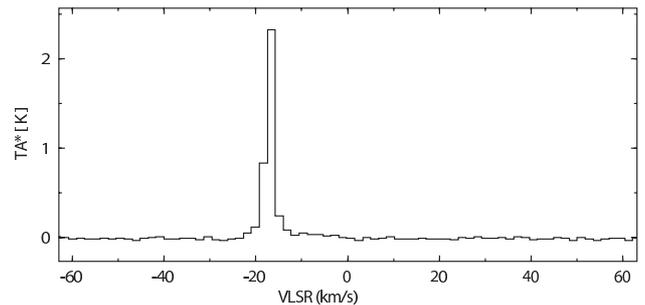}
\caption{Class I methanol maser emission spectrum from the CH$_3$OH (5-4) E transition at 84.521188 GHz.
\label{MethMaser}}

\end{center}

\end{figure}

The CH$_3$OH (5-4) E transition (84.521188 GHz) deviates from its expected excitation conditions as can be seen in the population diagram. This transition is known to exhibit maser behaviour \citep{Menten88}, therefore it seems that we have detected maser emission at this transition in S235AB (Figure~\ref{MethMaser}). The full line width of the emission is $11$ km s$^{-1}$ which is far too large to emanate from a single maser feature and thus indicates a likely blending of multiple features. The central velocity of this class I methanol maser is close to that of the molecular core, suggesting that it is not directly tracing the jet/outflow system and could instead be produced in the dense ambient gas somewhere near the outflow.

\null

\subsection{Optical depth using hyperfine emission lines}
%\subsubsection{C$_{2}$H}

Intensity ratios of the main line (F=2-1) to the  hyperfine satellite line (F=1-0) of the C$_{2}$H molecule can be used to derive the gas optical depth via:
\begin{eqnarray}
%Intensity ~ratio ~= \frac{T_{A}(1-0)}{T_{A}(2-1)} = \frac{1 - e^{-\tau}}{1 - e^{-a\tau}}
Line ~intensity ~ratio ~= \frac{1 - e^{-\tau}}{1 - e^{-a\tau}}
\end{eqnarray}

\noindent were $a$ is the intrinsic line intensity ratio determined by laboratory experiment. For C$_{2}$H (F=1-0):(F=2-1), $a=0.5$ in the optically thin limit \citep{Tucker74}. Therefore, using Equation 1 our observed line intensity ratio of $T_{A}(1-0) / T_{A}(2-1) =0.55$ gives $\tau_{C_2H} = 0.39$. This value is typical for young, massive star forming clumps \citep{Sakai10}.

For optically thin H$^{13}$CN the intrinsic intensity ratios of the hyperfine transitions (F=1-1):(F=2-1):(F=1-0) are 3:5:1 \citep{Irvine84}. Taking the transitions (F=2-1):(F=1-0)$= 5$, our observed intensity ratio of $T_{A}(F=2-1) / T_{A}(F=1-0) = 4.36$ gives $\tau_{H^{13}CN}=0.07$.

\subsection{Optical depth using isotopic abundance}
%Relative isotope abundances of C$^{32}$S/C$^{33}$S/C$^{34}$S revealed blah blah blah

We detected two isotopologues of two sulphur-bearing species; C$^{34}$S and CS, and $^{34}$SO and SO.
We estimate the optical depths of CS and SO using the isotopologue line intensity ratios, as was done in Equation 1, where $a$ is the intrinsic intensity ratio of $^{32}S/^{34}S = 22.5$ \citep{Kahane88}.
Our measured intensity ratio of $T_{A}(CS) / T_{A}(C^{34}S) = 4.9$ gives $\tau_{CS}=5.1$, and our intensity ratio of $T_{A}(^{34}SO) / T_{A}(SO) = 15.2$ gives $\tau_{SO}=0.9$.

%\subsection{Polarised emission}
%A transition of CH$_{3}$CHCN (9-8) exists at 85.715424 GHz [Lovas]. We detect this line at 85.719649 GHz corresponding to a velocity shift of -14.79 km s$^{-1}$, which generally agrees with the velocity of the parent cloud which is -17 km s$^{-1}$ \citep{Felli04} [check other lines in our survey].

%Intensities of this line were 5.17 times higher in the horizontal polarisation than that of the vertical polarisation. For validation we confirmed that the difference was not seen to vary with time, and that the other lines detected in this array had comparable intensities with regards to their horizontal and vertical polarisations. Furthermore, the emission covers several channels thus is not spurious. The vertical polarisation component is detected at 6$\sigma$ (0.0539 / 0.0091 K), while the horizontal component is 21$\sigma$ (0.2787 / 0.0131 K). Spectra of the vertical and horizontal polarisations are shown in Figures~\ref{VER} and \ref{HOR}.

%Polarisation of the CH$_{3}$CHCN (9-8) thermal line revealed blah blah blah

\subsection{Column densities and abundances}
To estimate column densities from individual molecular species we used the prescription outlined in \citet{Goldsmith99} for species with measurable optical depths (\emph{see} Sections 3.4 and 3.5) except for the column density of CH$_3$OH, which was obtained from the population diagram. For other molecules the statistical weights, upper energy levels, partition functions and Einstein coefficients for individual transitions were sourced from the Leiden Atomic and Molecular Database (LAMBDA) \citep{LAMBDA} and CDMS catalogues with an assumed excitation temperature of 18.75 K with the exception for SO for which we used the measured rotation temperature (\emph{see section 3.2}).

\citet{Saito07} estimate the molecular hydrogen column density of the molecular core in S235AB to be $10.1 \times 10^{22}$ cm$^{-2}$ based on their C$^{18}$O observations at comparable angular resolution. By comparing their results with the column densities measured in this work we estimated molecular abundances.
These, alongside the aforementioned physical gas parameters, are summarised in Table~\ref{table}.

% $X_{C_2H}= 3.5\times 10^{-8}$ 
% $X_{C_2H}= 5.7\times 10^{-10}$
% $X_{C_2H}= 6.6\times 10^{-8}$
% $X_{C_2H}= 9.7\times 10^{-9}$
% $X_{C_2H}= 7.8\times 10^{-8}$}

%[$10^{-10}$]
%350	
%5.7	
%660	
%97	
%7.8	

% H13CN http://www.mpia.de/MSF07/Posters/Zinchenko.pdf

\begin{table}[h!]
\caption{Summary of derived physical parameters.\label{table}}
\begin{center}
\small
\begin{tabular}{ccccccc}
\hline
Molecule&$\tau$& T${rot}$ & $N_{Tot}$ & Molecular \\
&& [K] & [cm$^{-2}$] & abundance \\ 
\hline
$C_2H$ 		& 0.39&	-		&3.50E+15&  $ 3.5\times 10^{-8}$ 	\\
$H^{13}CN$ 	& 0.07&	-		&5.80E+13&  $ 5.7\times 10^{-10}$	\\
$CS$ 		& 5.1 &	-		&6.63E+15&  $ 6.6\times 10^{-8}$	\\
$SO$		& 0.9 & $26\pm4$	&9.76E+14&  $ 9.7\times 10^{-9}$	\\
$CH_3 OH$ 	& $\ll$1&$10\pm1$	&7.85E+15&  $ 7.8\times 10^{-8}$	\\
\hline
\end{tabular}
%\begin{tablenotes}
%\end{tablenotes}
\end{center}
\end{table}

\section{Discussion and conclusions}

%\subsection{Jet driven high-velocity velocity wings}

Observations of S235AB by \citet{Sun12} detected SiO emission at a peak brightness consistent with our observations however velocity wings were not detectable at the 0.017 K rms sensitivity of their observations.
Our SiO spectrum (Figure~\ref{SP3}) exhibits wide velocity wings reaching up to $75$ km s$^{-1}$ from the core velocity. Our deeper observations confirm the presence of a fast molecular jet in S235AB. 
Discovery of a high-velocity jet in this source is of particular significance in light of recent water maser results of \citet{Burns15a}. Those authors find both blue-shifted and redshifted high velocity ($\sim$50 km s$^{-1}$ from the core velocity) maser features in a bipolar configuration matching the alignment of the slow ($< 5$ km s$^{-1}$ from the core velocity) NNW-SSE molecular outflow seen in HCO$^+$ and C$^{34}$S, reported by \citet{Felli04}. 

The velocity of the SiO gas ($\sim 75$ km s$^{-1}$) is much faster than that of the molecular outflow in S235AB making it likely that the SiO emission traces the primary molecular jet. Furthermore, as can be seen in Figure~\ref{SP3}, the SiO jet gas and water masers have similar terminal velocities, suggesting likely association - although this should be confirmed with mapping observation. 

Molecular outflows are thought to be produced as linear momentum from a protostellar jet is transferred into ambient gas around to the protostar. This gas becomes entrained at a slower velocity and at larger radii from the inner jet. 
The fast SiO gas in S235AB-MIR therefore indicates the presence of an entraining jet; the driving source for the water masers and molecular outflows seen in this region.

%Furthermore, \citet{Felli04} report observations of a slow  bipolar molecular outflow  which aligns with the water maser jet axis of \citet{Burns15a}. The slow molecular outflow therefore also represents entrained gas but at further proximity from the jet and therefore experiencing a lesser injection of momentum.

%The SiO jet in S235AB-MIR may represent the central link in the jet-outflow picture; the fast maser jet, the fast molecular jet, and the slow molecular outflow.

%The SiO spectrum is fit by two Gaussians: one for the main peak and one to parameterise the wings. The first Gaussian has a central velocity which is close to that of the molecular core, however the peak of the second Gaussian (the one dealing with the wings) is blueshifted by $-2.32$ km s$^{-1}$. Similarly, \citet{Burns15a} saw an asymmetry in favour of blueshifted rather than redshifted water masers with respect to the extremum velocities and number of detected maser features. Our interpretation is that of an asymmetric distribution of the outflow gas - perhaps due to stronger shocks in the blue shifted lobe, releasing more SiO into the gas phase and providing collisional energy to pump more water masers.

\citet{Burns15a} report evidence of possible jet rotation in their maser observations. Our confirmation of a molecular jet in S235AB-MIR introduce the possibility of investigating the spatial orientation of the jet via mapping observations - which may unearth further evidence of jet rotation.

%Small velocity wings are also seen in…

%\subsection{Physical parameters of the molecular gas}

Our survey investigated the physical properties of numerous molecular species in S235AB, which revealed a cold, dense gas environment with high opacity in several tracers - typical of the young cores in which MYSOs are embedded. Our survey results provide information on strengths and line widths of various gas tracers which will be useful for further follow-up observations.

We detected maser emission in the CH$_3$OH (5-4) E transition at 84.521188 GHz emanating from an ensemble of blended features. With regards to thermal CH$_3$OH emission at 97 GHz, we measured a low rotation temperature for the methanol gas, ($\sim10$ K), derived from the population diagram (Figure~\ref{MethMaser}).
In this respect S235AB conforms to the trend reported by \citet{Minier02} who find low rotation temperatures to be a systematic feature among their sample of massive star forming regions. They postulate that the 97 GHz methanol gas may trace cooler gas at the outer parts of the methanol cloud.

% is a systematic feature among regions of massive star formation and can be interpreted as originating in the cooler outer parts of the cloud \citep{Minier02}. 

With regards to molecular abundances we compare our results to \citet{Gerner14} who investigated a sample of 59 massive star forming regions categorised into different stages of evolution which, in order of ascending evolutionary stage are: infrared dark clouds, high mass protostellar objects, hot molecular cores and ultra compact H$_{\rm II}$ regions. 
A comparison with their results was possible for C$_2$H, SO, CS and CH$_3$OH which are common to ours and their works. For the cases of C$_2$H, SO, and CH$_3$OH our estimated abundances in S235AB are consistent with to the hot molecular core category of \citet{Gerner14} (\emph{see their} Figure 3), while CS matches somewhere between the hot molecular core and high mass protostellar object categories (\emph{see their} Figure 5). The abundance of H$^{13}$CN in S235AB is similar to those seen in another 5 well-studied massive star forming regions, the survey of which was reported in \citet{Zinchenko09}.

To conclude, the molecular abundances in S235AB are typical for a young massive star forming region. Its evolutionary stage close to that of a hot molecular core (\emph{this work}, \citealt{Felli04}) and preceding the formation of a HII region, as is indicated by the lack of centimeter emission \citep{Tofani95,Felli06}. Nonetheless it is a very active region as is evinced by the presence of slow molecular outflows \citep{Felli04}, a maser jet \citep{Burns15a}, and a fast molecular jet.

%This jet certainly merits further investigation.}

\section{Acknowledgements} 

We would like to thank staff at NRO for their generous support. We would also like to thank Takano Shuro for interesting discussion during the observing periods.

R.B. would like to acknowledge the Ministry of Education, Culture, Sports, Science and Technology (MEXT), Japan for financial support under the Monbukagakusho scholarship.

T. Hirota is supported by the MEXT/JSPS KAKENHI Grant Numbers 24684011, 25108005, and 15H03646

H.I. is also supported by the JSPS KAKENHI Grant Number 25610043.

K.M. is supported by a Grant-in-Aid from the JSPS Fellows 
and JSPS KAKENHI Grant numbers 24-6525 and 15K17613.

\footnotesize
\bibliographystyle{aa}
\bibliography{/Users/rossburns/Documents/LaTeX/Bib_Stuff/Kagoshima.bib}

\end{document}